\documentclass[%
 reprint,
 amsmath,amssymb,
prb,
]{revtex4-2}

\usepackage{graphicx}
\usepackage{dcolumn}
\usepackage{bm}
\usepackage{xcolor,soul,framed} 
\usepackage{subfigure}
\usepackage{fancyhdr}
\usepackage{booktabs}
\usepackage{tabularx}
\usepackage{comment}
\usepackage{amsmath} 
\usepackage{blindtext}
\renewcommand{\figurename}{Figure}

\newcommand{\Fig}[1]{Figure\,{\ref{#1}}}

\newcommand{\E}{\mathrm{e}}
\newcommand{\jj}{\mathrm{j}}

\begin{document}


\title{{Transient States to Control the Electromagnetic Response of\\ Space-time Dispersive Media}}
\author{Pablo H. Zapata-Cano$^1$, Salvador Moreno-Rodríguez$^2$, Stamatios Amanatiadis$^1$, Antonio Alex-Amor$^{3}$, Zaharias D. Zaharis$^{1}$, Carlos Molero$^{2}$}

\affiliation{\vspace{0.1cm}$^1$School of Electrical and Computer Engineering, Aristotle University of Thessaloniki, 54124 Thessaloniki, Greece \vspace{0.1cm}\\ 
$^2$Department of Signal Theory, Telematics and Communications, Research Centre for Information and Communication Technologies (CITIC-UGR), Universidad de Granada, 18071 Granada, Spain \vspace{0.1cm}\\ 
$^{3}$Department of Electrical and Systems Engineering, University of Pennsylvania,  Philadelphia, Pennsylvania 19104, United States
}%
\email{Corresponding author: pablozapata@auth.gr}

\begin{abstract}
In this paper, we study the dynamic formation of transients when plane waves impinge on a dispersive slab that abruptly changes its electrical properties in time. The time-varying slab alternates between air and metal-like states, whose frequency dispersion is described by the Drude model. It is shown how the physics of this complex system can be well described with the joint combination of two terms: one associated with temporal refractions and the other associated with spatial refractions. To test the validity of the approach, some analytical results are compared with a self-implemented finite-difference time-domain (FDTD) method. Results show how the transients that occurred after the abrupt temporal changes can shape the overall steady-state response of the space-time system. In fact, far from always being detrimental, these transient states can be conveniently used to perform frequency conversion or to amplify/attenuate the electromagnetic fields.  
\end{abstract}

\maketitle


\section{Introduction}

The electrodynamics of time-varying media have been an object of study since the middle of the last century. The pioneering works of Morgenthaler, Fante, Felsen, and Whitman \cite{morgenthaler1958velocity, Felsen70, Fante71, Fante73} laid the theoretical foundations for the analysis of this type of complex media. However, the difficulty of applying the concepts developed therein to structures that could be feasible in practice left this line of research somewhat aside.

With the vast development of metamaterials, the research field of time-varying media has regained the attention of both scientific and engineering communities over the past few years \cite{caloz2019spacetime, galiffi2022photonics}. The so-called \emph{space-time metamaterials} have revolutionized and extended the former vision of light-matter interactions \cite{Tirole2023, Pachecho2021, Kort-Kamp2021, stefanini2022}, and numerous microwave, photonics and optical applications are emerging from their study. To just name a few of these: Doppler cloaking \cite{Fang2023}, frequency conversion and mixing \cite{taravati2018aperiodic, Amra24}, beamforming \cite{taravati2020full}, non-reciprocal antenna response \cite{zang2019nonreciprocal}, Faraday rotation \cite{Huan2023} or magnet-free circulator design \cite{Mock2019}.

In our previous works \cite{alex2023diffraction,  moreno2023time, Moreno24eucap, alex2023analysis,   Moreno24spacetime}, we have considered scenarios where space-time-modulated metamaterials periodically transition between air and metal states. In these approximations, metallic elements were ideally treated as perfect electric conductors (PEC), while air was simply modeled as a dielectric of unitary relative permittivity and permittivity.  This consideration, although approximate, led to analytical frameworks that revealed interesting physical properties of the aforementioned devices. 

In fact, the modeling of metallic elements as PEC is quite convenient, since it imposes the nullity of tangential electric fields at the spatial interfaces, thus greatly reducing the complexity of the space-time problem. The consideration of metallic elements as PEC is broadly applied in radio and microwave regimes, as it generally leads to accurate results. However, this may not be the case for infrared light and, with all certainty, visible light and light of greater frequencies. Waves may become partially transparent inside the metal above a certain frequency limit \cite{OpticalBook}. 

\emph{Frequency dispersion}, the fact that the electrical and optical properties of a certain material change with frequency is indeed expected to play a relevant role in the air-metal transitions. As detailed in some reference works \cite{caloz2019spacetime, Gratus_2021, Monticone22}, the space-time boundary conditions of the electromagnetic fields are significantly affected by the presence or absence of frequency dispersion, and so is the temporal response of the system. As an example, electric-field and polarizability vectors are continuous (discontinuous) across temporal interfaces in frequency-dispersive (nondispersive) materials.

Naturally, real-world materials are always, to a greater or lesser extent, frequency dispersive. Nonetheless, the vast majority of theoretical efforts have been put into successfully describing the foundations of time-varying nondispersive scenarios due to the simpler associated analysis  \cite{morgenthaler1958velocity, Fante73, zurita2009reflection, Xiao:14, taravati2019generalized, huidobro2021homogenization, GarciaVidal24} (except for a few, among which we highlight the recent contributions \cite{Solis21, Mirmoosa2022, Amra24}).\\

It can be straightforwardly shown that abrupt temporal changes that occur in non-dispersive scenarios produce instantaneous field variations with no associated transients. On the other hand, scenarios with frequency-dispersive materials create intermediate transient states that connect with the final steady-state response of the system. This is analogous to the transient response that takes place in the charge/discharge of a capacitor. 

The use of ideal dispersionless air and PEC conditions in our previous works \cite{alex2023diffraction,  moreno2023time, Moreno24eucap, alex2023analysis,   Moreno24spacetime} limited the study of transients in air-to-metal and metal-to-air transitions. In a PEC, free charges react instantly (zero time) to an external excitation, so transients no longer exist. In this paper, we analyze the formation of transients due to air-Drude transitions and their impact on the response of metallic metamaterials driven by space-time modulations. The frequency-dispersive and lossy nature of metals is represented here by means of the Drude model, which stands as an accurate approximation for studying the dispersive behavior of materials whose electromagnetic response is governed by free carriers. It is not only applicable to classic plasmonic metals such as gold, silver, copper, or aluminum, but it also includes many other materials with a high free carrier density. In this scenario, we find materials such as graphene, transparent conductive oxides (TCOs), transition metal dichalcogenides (TMDs), highly-doped semiconductors like silicon (Si) or gallium arsenide (GaAs), that allow tuning their free carriers with external electrical or optical stimulation \cite{abdelraouf2022recent}. In addition, metamaterials and structured platforms whose effective parameters follow a Drude-like dispersive response can be also described with the present analytical framework, provided that they show a non-magnetic effective response and are not resonant in the working frequency range.

Frequency dispersion may shape or even completely modify the physical behavior of the spatiotemporal system. In the following, the Drude model is integrated into first-principle calculations to accurately predict the transient responses of a set of metal-based spatiotemporal metamaterials. In contrast to other works in which a ``temporal slab" is considered on its own (the whole medium changes its material properties at every transition), here we propose a systematic analytical approach in which the effect of the temporal transient is added to the repercussion of a spatial component derived from the continuous impingement of an incident plane wave that travels in the free space and encounters the time-modulated, frequency-dispersive, material slab. Moreover, it is showcased that transients, far from being detrimental, add richness to the problem and can help to tune the electromagnetic response of the space-time device. In fact, controlled transient states can enable an efficient way for a number of engineering applications such as frequency conversion, field amplification/attenuation, or scattering control, among others.



\section{Air-to-Drude Transition}

\begin{figure}[!t]
	\centering
    \includegraphics[width= 1\columnwidth]{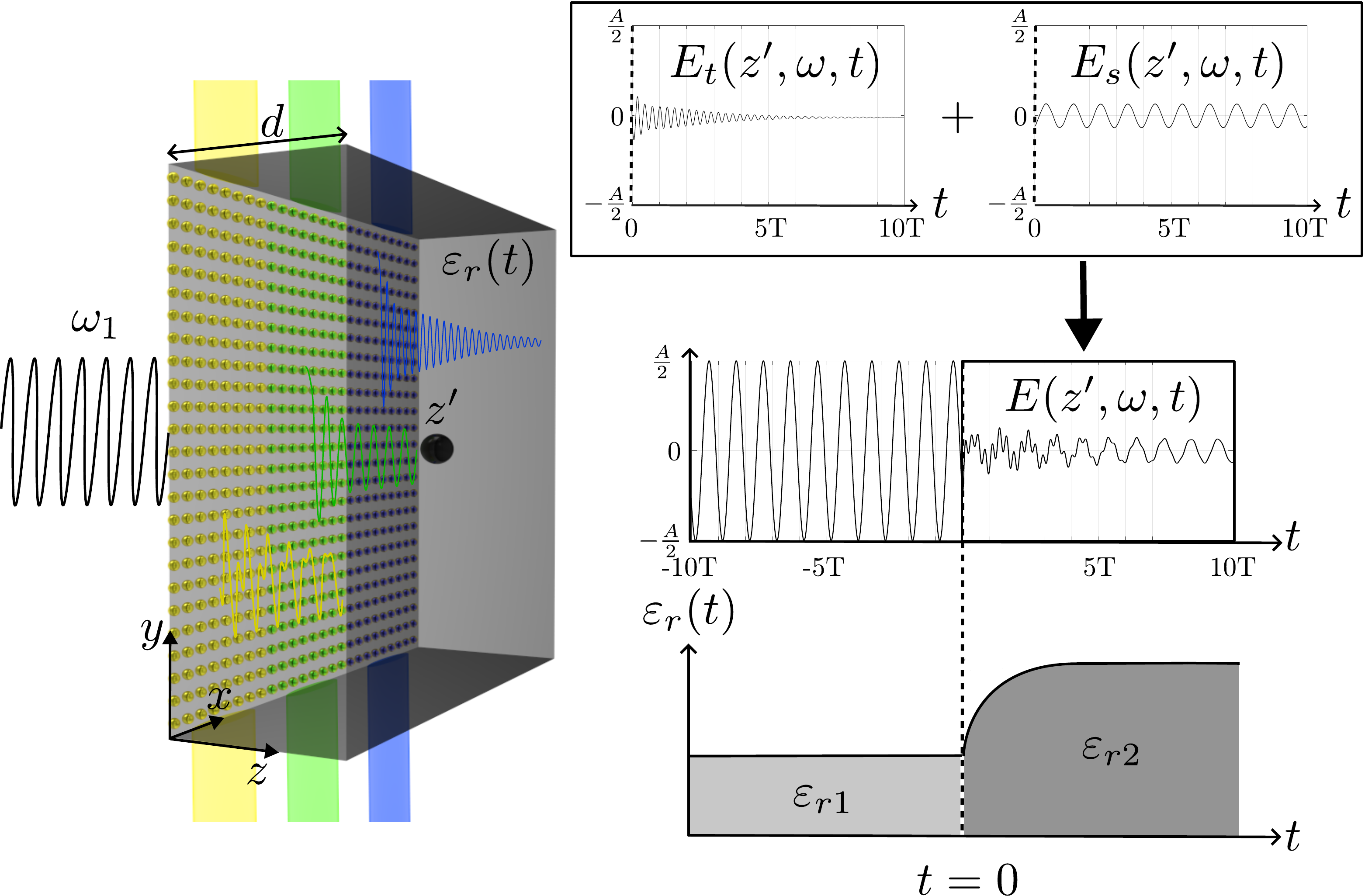}
	\caption{An incident wave of frequency $\omega_1$ impinges on a dispersive time-varying element that alternates between air and Drude-like states. The transient electric field at position $z'$ is greatly affected by the parameters of the Drude material, reconfigurable with lasers (colored lights) or external bias, and the temporal modulation. Transient states can be used to tune the overall electromagnetic response of the device.}
	\label{sketch}
\end{figure}

In this section, we analyze the transients occurring in the spatiotemporal system sketched in Figure \ref{sketch}. An incident plane wave of frequency $\omega_1$ propagating in air impinges on a homogeneous, isotropic, and frequency-dispersive slab of thickness $d$. The slab, located at $0\leq z \leq d$ and infinitely extended along the $x$ and $y$ axes, abruptly transitions at the instant $t=0$ its electrical properties from air to a Drude-like material described by the Drude model. The Drude-like state can represent a real-world non-magnetic metal, a lossy and dispersive dielectric, or a metamaterial with non-magnetic dispersive effective parameters. Thus, for $t<0$, the whole space is filled with air. For $t\geq 0$, the region between $0\leq z \leq d$ changes its electrical properties to a Drude-like state.

Air is modeled here as a dielectric with parameters $\varepsilon_r = \mu_r = 1$. The electrical properties of the non-magnetic Drude-like state are represented by $\mu_r = 1$ and a relative permittivity term 
\begin{equation} \label{epsilon_w}
\varepsilon_r(\omega) = 1 - \frac{\omega_p^2}{\omega^2 - j\omega\gamma }\, ,
\end{equation}
where $\omega_p$ and $\gamma$ are the plasma and damping frequencies, respectively \cite{Ordal83}.

The plasma frequency $\omega_p$, typically of the order of $\omega_p \sim 10^{16}$ rad/s in good conductors such as silver, gold, copper, or aluminum, marks the transition between the Drude medium acting as a mainly-reflecting ($\omega \ll \omega_p$), or mainly-transparent ($\omega \gg \omega_p$) element. The plasma frequency may be computed as $\omega_p = \sqrt{Nq_e^2 / (\varepsilon_0 m_e)}$, where $N$ is the free-electron volumetric density, $q_e$ and $m_e$ are the electron's charge and mass, respectively. The damping frequency or damping constant $\gamma$, of the order of $\gamma \sim 10^{14}$ Hz in good conductors, is a loss term that accounts for energy loss in free-electron collisions. It can be modeled as $\gamma = q_e/(\mu_e m_e) \,$, where $\mu_e$ represents the carrier mobility \cite{lang2020tunable}.

In essence, the parameters $\omega_p$ and $\gamma$ will shape the transient response of the system after the temporal change occurs at $t=0$. Interestingly,  $\omega_p$ and $\gamma$ can be externally tuned in the laboratory in some Drude-like materials. This enables a practical way of controlling the transient states in the present space-time metadevice. 
In this topic, TCOs have attracted great attention due to their facilities to modify the plasma frequency. Specifically, TCOs like Al-doped ZnO (AZO) \cite{george2016localized} or indium tin oxide (ITO) \cite{feigenbaum2010unity}, allow to tune their free-carrier density through external electric stimulation \cite{huang2016gate,george2017electrically,kafaie2018dual,park2021all} or laser excitation \cite{guo2016ultrafast}. All these works are focused on the tunning of the carrier density to alter the complex relative permittivity of the materials. However, the application of strain engineering \cite{sun2024nanoimprint}, has enabled to improve the carrier mobility in TMDs as MoS$_2$ or WSe$_2$ \cite{7182773}. In fact, it has been possible to modify the damping frequency without altering the free-carrier density \cite{ng2022improving}. Moreover, with the aim to allow ultrafast modulations of the refraction index, some alternatives based on GaAs \cite{vabishchevich2021ultrafast,karl2020frequency}, high-purity graphene \cite{ni2016ultrafast} or germanium (Ge) \cite{he2021ultrafast} have spread recently. These works open the door to independently and rapidly tune these two parameters related to the plasma frequency and damping frequency.

When the slab abruptly changes over time its electrical properties from air to Drude, the plane wave \emph{trapped} inside the slab has the following temporary evolution in terms of its associated transverse electric field:
\begin{equation} \label{E}
    E(z,t) \approx E_\text{t}(z, t) + E_\text{s}(z, t)
\end{equation}
where $E_{\text{t}} (z, t)$ is the temporal refraction term and $E_{\text{s}}(z, t)$ is the spatial refraction term. A similar rationale was used in \cite{Fante73} in the absence of frequency dispersion, with the simplifications that this fact entails. The temporal refraction term $ E_\text{t}(z,t)$ is mainly governed by the \emph{time-interface} problem, in which the plane wave suffers a temporal refraction caused by the \emph{instantaneous} change of the medium properties at $t=0$.  On the other hand, the spatial refraction response $ E_\text{s}(z,t)$ is a stationary term governed by the \emph{space-interface} problem. This term is due to the presence, and continued spatial refraction at the boundaries $z=0$ and $z=d$, of the incident plane wave on the Drude material.

The analysis proposed via eq. \eqref{E} is a simplified strategy that gives accurate results under certain conditions. The main restriction to obtain accurate results is that the thickness of the slab $d$ is not too thin compared to the operation wavelength. This restriction is particularly relevant for the following reason: The fact of considering a slab that is not too thin precisely allows us to subdivide the complex general space-time-varying problem into its time-only and space-only subproblems governed by the temporal and spatial refraction terms, respectively. As will be discussed below, the abrupt change of the material properties at $t=0$ generates two forward and backward waves that propagate in the slab. These two newly generated waves have associated complex frequencies, with their imaginary parts relating to absorption in the Drude material. Therefore, both forward and backward waves rapidly attenuate over time. If the slab is thick enough, then the waves mostly attenuate before reaching the spatial boundaries, preventing secondary spatial refractions. In this scenario, the physics of the problem is well captured by the approximation given in eq. \eqref{E}.

The contribution of both temporal and spatial refraction terms approximately describes the spatiotemporal system in a simple manner. This way of analysis is advantageous since both time and space light-matter interactions can be independently studied, leading to a better comprehension of the whole scenario.  Naturally, the sum in eq. \eqref{E} is constrained by causality conditions. Right before the temporal change ($t=0^-$), the incident plane wave occupies the whole space inside the region of the slab, $0\leq z \leq d$. Thus, right after the temporal change ($t=0^+$), two forward and backward waves generate
and travel simultaneously at every point $z$ inside the slab. This will take account of the temporal refraction term $E_\text{t}(z, t)$. To all of the above, we have to add the contribution of spatial refraction term $E_\text{s}(z, t)$. However, the spatial refraction term has a wave associated with it that starts to propagate from $z=0$. As a result, the spatial refraction term takes a time $t_0=z_0/v$, with $v$ being the velocity in the Drude-like material, to move from $z=0$ to the position $z=z_0$. This fact entails that the spatial refraction term cannot be added up to the temporal refraction term in a region where the wave in question has not arrived yet.

\subsection{Temporal Refraction Term $E_{\text{t}}(z, t)$}
\label{sec:temporal_refraction_term}

The transient behavior of the electric field is obtained after imposing the boundary conditions to be satisfied by the electromagnetic fields in media with temporal variation \cite{Monticone22}. It is well known that vectors $\mathbf{D}$ and $\mathbf{B}$ are continuous in a medium whose electrical properties vary in time. Thus, $\mathbf{D}(\mathbf{r}, t = 0^{-}) = \mathbf{D}(\mathbf{r}, t = 0^{+})$ and $\mathbf{B}(\mathbf{r}, t = 0^{-}) = \mathbf{B}(\mathbf{r}, t = 0^{+})$,
where it has been assumed that the medium changes its properties at $t = 0$. In fact, in the absence of appreciable magnetism ($\mu = \mu_0$), the continuity of vector $\mathbf{B}$ leads to the the condition $\mathbf{H}(\mathbf{r}, t = 0^{-}) = \mathbf{H}(\mathbf{r}, t = 0^{+})$.

Additionally, the electric susceptibilities of the air and Drude media are $\chi(t)=0$ and $\chi(t) = \frac{\omega_{\text{p}}^{2}}{\gamma} \left(1 - \text{e}^{- \gamma t}\right) U(t)$, respectively \cite{Taflove}, with $U(t)$ representing the unit step function. It can be noted that the electric susceptibility $\chi(t)$ is continuous and nulls at $t=0$. Recalling that $\mathbf{D} = \varepsilon_0 \mathbf{E} +  \mathbf{P}$ is continuous at $t=0$, with $\mathbf{P} = \varepsilon_0 \chi * \mathbf{E}$ (``$*$" represents the temporal convolution), it can be inferred the continuity of the polarization charge and electric field too; namely,  $\mathbf{P}(\mathbf{r}, t = 0^{-}) = \mathbf{P}(\mathbf{r}, t = 0^{+})$  and  $\mathbf{E}(\mathbf{r}, t = 0^{-}) = \mathbf{E}(\mathbf{r}, t = 0^{+})$. The continuity of $\mathbf{P}$ and $\mathbf{E}$ should be attributed to the presence of frequency dispersion, a fact that is also stated in \cite{Solis21, Monticone22} for a Lorentzian dispersive medium. The continuity of the polarization charge and electric field vectors is not satisfied in nondispersive scenarios with abrupt temporal changes \cite{morgenthaler1958velocity, caloz2019spacetime}. 

The boundary conditions also force the wavevector of the wave to be continuous across the time interface:
\begin{equation}\label{cont_k}
    k_{1} = \frac{\omega_1}{c} =  k_{2} =\frac{\omega_2}{c} \sqrt{1 - \frac{\omega_p^2}{\omega_2^2 - j\omega_2\gamma }}\, ,
\end{equation}
where $k_{1} = k(t = 0^{-})$ and $k_{2} = k(t = 0^{+})$.
Notice that eq. \eqref{cont_k} imposes a frequency jump from $\omega_1$ to $\omega_2$.

It can be demonstrated that there exist three solutions for eq. \eqref{cont_k}. Unfortunately, the mathematical form of the three mentioned solutions is rather cumbersome. However, under some approximations, the expression for new frequency $\omega_2$ after the temporal change is significantly simplified. We will focus on the scenarios in which the Drude material behaves like a metal (metal-like regime) and a dielectric (dielectric-like regime). The metal-like regime approximation will hold as long as the Drude material behaves as a good-conducting material. It can be shown that the metal-like approximation works for a wide range of frequencies, from DC to approximately $0.1\omega_p \sim 10^{15}$ rad/s. On the other hand, the dielectric-like approximation will hold as long as the Drude material behaves as a dielectric with a small, but not necessarily negligible, loss term. That is the case if $\omega \gg \gamma$. See the supplementary material \cite{SupplementaryMaterial} for further information on the considered approximated regimes.

In the metal-like approximation ($\omega_1 \rightarrow 0$, in practice, $\omega_1 \ll \gamma, \omega_p$), the solutions for the new frequency $\omega_2$ after the temporal change occurred at $t=0$ are \linebreak $\omega_2 \approx \frac{1}{2} \left(\pm \sqrt{4\omega_p^2 - \gamma^2} +\jj\gamma \right)$. Note that the two solutions are complex-valued. The real part of $\omega_2$ represents the oscillation frequency of the newly generated waves, while the imaginary part indicates that the waves attenuate over time proportionally to $\E^{\frac{\gamma}{2}t}$. In the usual scenario for good-conducting metals such as copper, gold, or aluminum, the plasma frequency is normally much larger than the damping frequency, i.e., $\omega_p \gg \gamma$. Therefore, the new frequencies $\omega_2$ can be further approximated to $\omega_2 \approx \pm \omega_p + \jj \gamma/2$. This fact denotes that the complex frequencies $\omega_2$ generated after the abrupt temporal change vibrate with a frequency close to the plasma frequency in the metal-like regime.


In the dielectric-like regime, loss terms will not be relevant and it can be assumed that $\gamma \rightarrow 0$ (in practice, $\omega_1 \gg \gamma$). Under dielectric-like conditions, the two solutions for $\omega_2$ are $\omega_2 \approx \pm \sqrt{\omega_1^2 + \omega_p^2}$. Note that, in this case, the solutions of $\omega_2$ are real-valued, as a difference with respect to the metal-like regime.  This is due to the fact that $\gamma \rightarrow 0$ (loss terms are negligible). The reader is referred to the supplementary material \cite{SupplementaryMaterial} for further information on the frequency conversion and complex frequencies.  

In both metal-like and dielectric-like regimes, the plus (+) and minus (-) signs in the solutions of $\omega_2$ represent forward and backward waves propagating in space, respectively. Forward and backward propagation are referred to the $+z$ and $-z$ directions, respectively. Thus, the overall physical behavior of the temporal refraction term could be summarized as follows. The medium in which the original incident wave of frequency $\omega_1$ is propagating suffers an abrupt change of its electrical properties at $t=0$. The continuity of the wavenumber $k$ and fields $\mathbf{D}$ and $\mathbf{B}$ over the temporal discontinuity provokes that the original incident wave is split into two waves that travel forward and backward in space. In a general scenario, the frequencies after the temporal change are complex, thus showing that the fields not only vibrate but also attenuate over time. In the particular case of working in the dielectric-like regime, losses are neglected and the new frequencies $\omega_2$ are real-valued. 

The evolution of the electric field associated with the temporal refraction term can be extracted by solving the Helmholtz equation. In the source-free case, it can be expressed in the following form:
\begin{equation} \label{Helmholtz}
    \nabla \times \nabla \times \mathbf{E}_{\text{t}}(\mathbf{r}, t) = -\mu_0 \varepsilon_0 \frac{\partial^2 \mathbf{E}_{\text{t}}(\mathbf{r}, t)}{\partial t^2 } - \mu_0 \frac{\partial^2 \mathbf{P}(\mathbf{r}, t)}{\partial t^2 }
\end{equation}
Since we are analyzing a (1+1)-D problem, the vector notation can be dropped. Additionally, it should be noted that $\nabla \times \nabla \times \mathbf{E}_{\text{t}}(\mathbf{r}, t) = k^2 E_{\text{t}}(z,t)$. Thus, eq. \eqref{Helmholtz} is simplified to
\begin{equation}
    k^2E_{\text{t}}(z,t) =  -\mu_0 \varepsilon_0 \frac{\partial^2 E_{\text{t}}(z, t)}{\partial t^2 } - \mu_0 \frac{\partial^2 P(z, t)}{\partial t^2 }
\end{equation}
The differential equation in \eqref{Helmholtz} can be solved by employing the Laplace-transform formalism, as carried out in \cite{Solis21} for a time-varying Lorentz-dispersive media. The application of the Laplace transform to both members of \eqref{Helmholtz} leads to
\begin{multline} \label{helm1}
    k^2 E_{\text{t}}(z,s) = \\ -\mu_0 \varepsilon_0 \Bigg[s^2 E_{\text{t}}(z,s) - s E_{\text{t}}(z, t = 0^{-})   - \frac{\partial E_{\text{t}}(z, t = 0^{-})}{ \partial t} \Bigg]  \\- \mu_0 \left(s^2 P(z,s) - s P(z, t = 0^{-}) - \frac{\partial P(z, t = 0^{-})}{ \partial t} \right)
\end{multline}
The original plane wave, before the change of the medium permittivity ($t<0$), was a regular plane wave described by $E_{\text{t}}(z, t) = A\,  \text{e}^{\text{j} \omega_1t}\, \text{e}^{-\text{j}k z} $, with $A$ being an arbitrary complex-valued constant. Thus 
\begin{align} \label{ini1}
E_{\text{t}}(z, t = 0^{-}) &= A\,\text{e}^{-\text{j} k z} \\
\frac{\partial E_{\text{t}}(z, t = 0^{-})}{\partial t } &= \text{j} \omega_1 A\,  \text{e}^{-\text{j} k z}\,.
\end{align}
Furthermore, the input media before the change is air, so the polarizability must satisfy
\begin{equation} \label{ini3}
    P(z, t = 0^{-}) =   0 = \frac{\partial P(z, t = 0^{-})}{\partial t }
\end{equation}
The insertion of the initial conditions \eqref{ini1}-\eqref{ini3} in the Helmholtz equation \eqref{helm1}, together with the use of the relation $P(z,s) = \varepsilon_{0}\chi(s) E_{\text{t}}(z, s)$, lead to 
\begin{equation}
E_{\text{t}}(z, s) \Big[k^2 c^2 + s^2 \big(1 + \chi(s) \big)\Big] = (s + \text{j}\omega_1) A\text{e}^{-j k z}\,.
\end{equation}
With the knowledge of $k^{2} = \omega_1^2 /c^2$ and $1+\chi(s) = \varepsilon_r(s)$, $\varepsilon_r(s)$ defined in eq. \eqref{epsilon_w} for $s=\jj\omega$, the temporal refraction term admits to be represented in the Laplace domain as
\begin{equation}\label{E_Laplace}
E_{\text{t}}(z, s) = A\text{e}^{-\text{j}k z}\, \frac{(s + \text{j}\omega_1) (s+\gamma)}{s^2 (s+\gamma) + s(\omega_p^2 + \omega_1^2) +\omega_1^2 \gamma} \,.
\end{equation}

The application of the initial and final value theorems \cite{BookLaplace} allows us to directly estimate the temporal response from the Laplace domain at the instants $t\rightarrow 0^+$ and $t\rightarrow \infty$. The initial value theorem states that \linebreak $E_t(z, t\rightarrow 0^+) = \lim_{s\rightarrow \infty} s E_t(z, s) = A\text{e}^{-\text{j}k z} $. Therefore, the electric field is continuous at the temporal jump, $ E_t(z, t\rightarrow 0^-) =  E_t(z, t\rightarrow 0^+)$. The final value theorem states that $E_t(z, t\rightarrow \infty) = \lim_{s\rightarrow 0} s E_t(z, s) = 0$, showing that the electric field is bounded indeed. 

We can recover the temporal evolution $E(z,t)$ by applying the inverse Laplace transform to eq. \eqref{E_Laplace}. Unfortunately, the expression of eq. \eqref{E_Laplace} lacks of analytical inverse transform in a general case, so it has to be computed numerically. Nonetheless, analytical expressions can be obtained for the metal-like and dielectric-like regimes.  For the metal-like regime ($\omega_1 \rightarrow 0$), the application of the inverse Laplace transform to eq. \eqref{E_Laplace} leads to the temporal refraction term
\begin{multline} \label{Et_low}
    E_t(z,t) \approx \mathrm{e}^{-\jj k z} \Big[E_{21} \mathrm{e}^{+\jj\left(  \frac{1}{2} \sqrt{4\omega_p^2 - \gamma^2}\right) t} \mathrm{e}^{ - \frac{\gamma }{2} t }  \\  + E_{22} \mathrm{e}^{-\jj\left(  \frac{1}{2} \sqrt{4\omega_p^2 - \gamma^2}\right) t} \mathrm{e}^{ - \frac{\gamma }{2} t } \Big]\, , 
\end{multline}
with $E_{21} = A/2\, \Big(1 - \jj \gamma / \sqrt{4\omega_p^2 - \gamma^2}\Big) \stackrel{\gamma \ll \omega_p }{\approx} A/2$ and \linebreak $E_{22} = A/2\, \Big(1 + \jj \gamma / \sqrt{4\omega_p^2 - \gamma^2} \Big) \stackrel{\gamma \ll \omega_p }{\approx} A/2$. 

For the dielectric-like regime ($\gamma \rightarrow 0$), a similar procedure leads to the temporal refraction term
\begin{multline} \label{Et_high}
    E_t(z,t) \approx \mathrm{e}^{-\jj k z} \Big[E_{21} \mathrm{e}^{+\jj\left(\sqrt{\omega_p^2 + \omega_1^2}\right) t}   \\  + E_{22} \mathrm{e}^{-\jj\left(\sqrt{\omega_p^2 + \omega_1^2}\right) t} \Big]\, , 
\end{multline}
with $E_{21} = A/2\, \Big(1 +  \omega_1 / \sqrt{\omega_p^2 + \omega_1^2}\Big)$ and \linebreak $E_{22} = A/2\, \Big(1 -  \omega_1 / \sqrt{\omega_p^2 + \omega_1^2}\Big)$.

In both regimes, the terms $E_{21}$ and $E_{22}$ represent the electric-field amplitude of the forward and backward waves, respectively, generated in the Drude material after the temporal discontinuity at $t=0$. As seen, the frequencies of the forward and backward waves coincide with the values of $\omega_2$ previously estimated by applying the continuity of $k$, and so does the exponential decay $\mathrm{exp}(-t\gamma / 2 )$. In good conducting metals ($\gamma \sim 10^{14}$ Hz), the temporal refraction term $E_t(z,t)$ rapidly attenuates, in just a few tens of femtoseconds, unless the incident frequency $\omega_1$ is extremely high.  Moreover, note that the electric-field amplitude of the DC wave solution ($\omega_2 = 0$) vanishes, as previously theorized.

\subsection{Spatial Refraction Term $E_{s}(t, z)$}

The spatial refraction term $E_{s}(t, z)$
is a steady-state component that takes into account the refraction of the incident wave on the spatial interfaces. Note that, after the abrupt temporal change of the dielectric properties that occurred at $t=0$, the incident wave keeps impinging on the slab. The interaction of the incident wave, of frequency $\omega_1$, with the spatial interface causes part of it to be reflected back to the air medium ($z<0$) and part to be transmitted to the Drude material ($z\geq 0$). The scenario is illustrated in Figure \ref{sketch}. 

As a difference with respect to the temporal refraction term, the boundary that separates air and Drude at $z=0$ is of spatial nature. In a spatial interface, the frequency must be continuous ($\omega_1=\omega_2$) while the wavenumber changes ($k_1 \neq k_2$). Note that the situation was just the opposite at the time interface $t=0$ regulating the temporal refraction term. Under the continuity of the frequency at the spatial interface, it can be inferred that the complex wavenumber $k_2 = (\omega_1 / c) \sqrt{\varepsilon_{\text{r}}(\omega_1)} = k$, with $\varepsilon_{\text{r}}(\omega_1)$ being computed by inserting $\omega_1$ in eq. \eqref{epsilon_w}. 

The evaluation of the spatial refraction field inside the slab can be easily performed by applying the theory of small reflections \cite{pozar}. The refracted \emph{standing} waves in the frequency steady regime are actually formed by the contribution of multiple reflections/refractions inside the slab (transient). When the considered material is lossy, which is the case of the Drude medium, the field at a given spatial point inside the slab is well described by 
\begin{equation} \label{Es}
E_s(z,t) = A T_{\text{a-D}}\, \text{e}^{\text{j} \omega_1 t}\,   \text{e}^{-\text{j} k z}  \, ,
\end{equation}

\noindent where $T_{\text{a-D}} = 2 / \big[1 + \sqrt{\varepsilon_{\text{r}}(\omega_1)} \big]$
is the transmission coefficient from air to Drude media. Note that the transmission coefficient can be greater than one in scenarios where $|\varepsilon_{\text{r}}(\omega_1)| < 1 $, e.g., when the incident frequency is close to the plasma frequency. This could lead to electric-field amplification.  

The expression for $E_s$ in eq. \eqref{Es} only considers the first spatial refraction at $z=0$. Reflections coming from the second spatial interface, $z=d$, are of much smaller amplitude and can be neglected, provided that the Drude medium is partially lossy and that the considered slab is not too thin. In the case of working with really thin slabs, secondary reflections must be included to achieve accurate results.


\begin{figure*}[!t]
	\centering
    \subfigure[]{\includegraphics[width= 0.8\columnwidth]{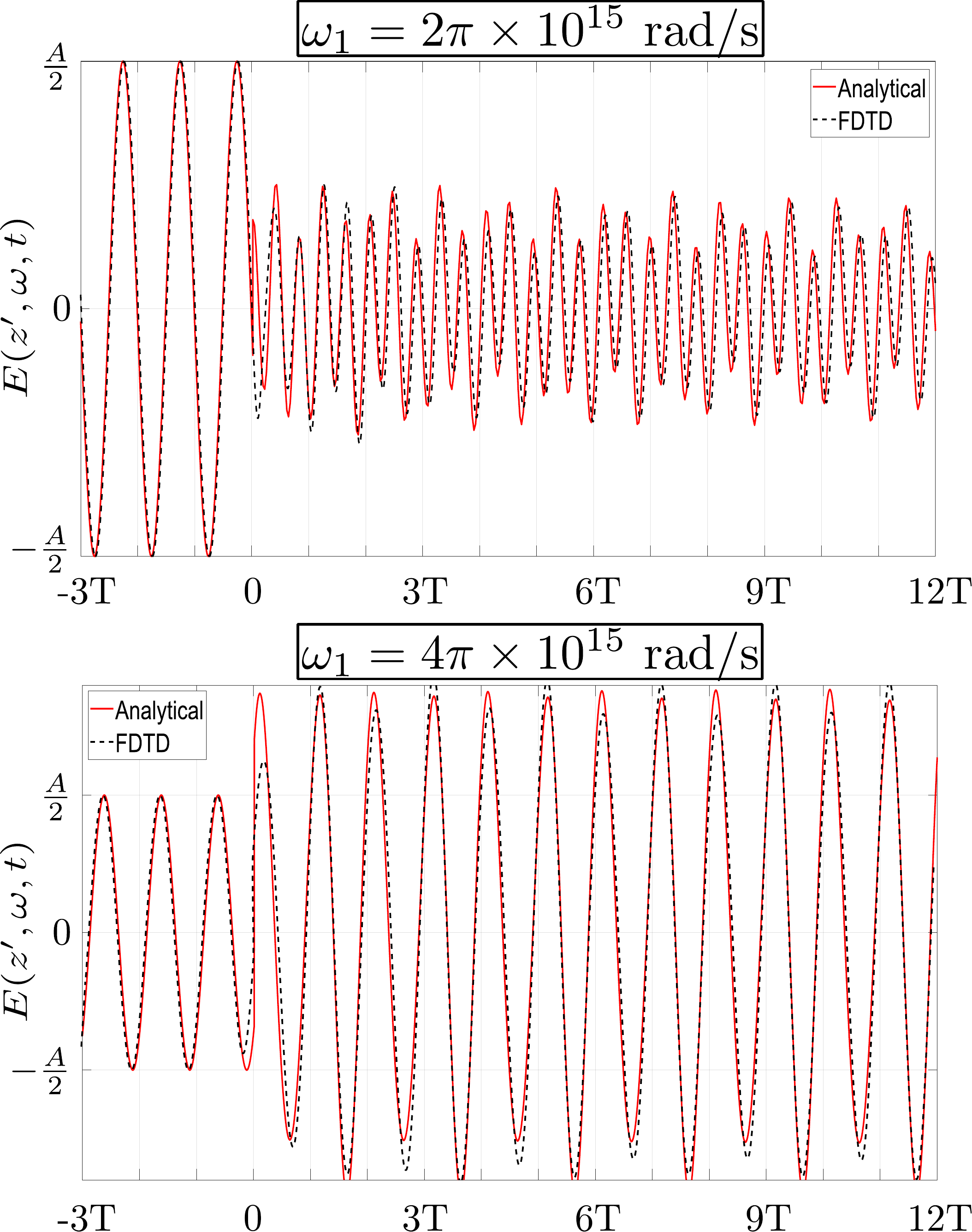}}
    \subfigure[]{\includegraphics[width= 0.8\columnwidth]{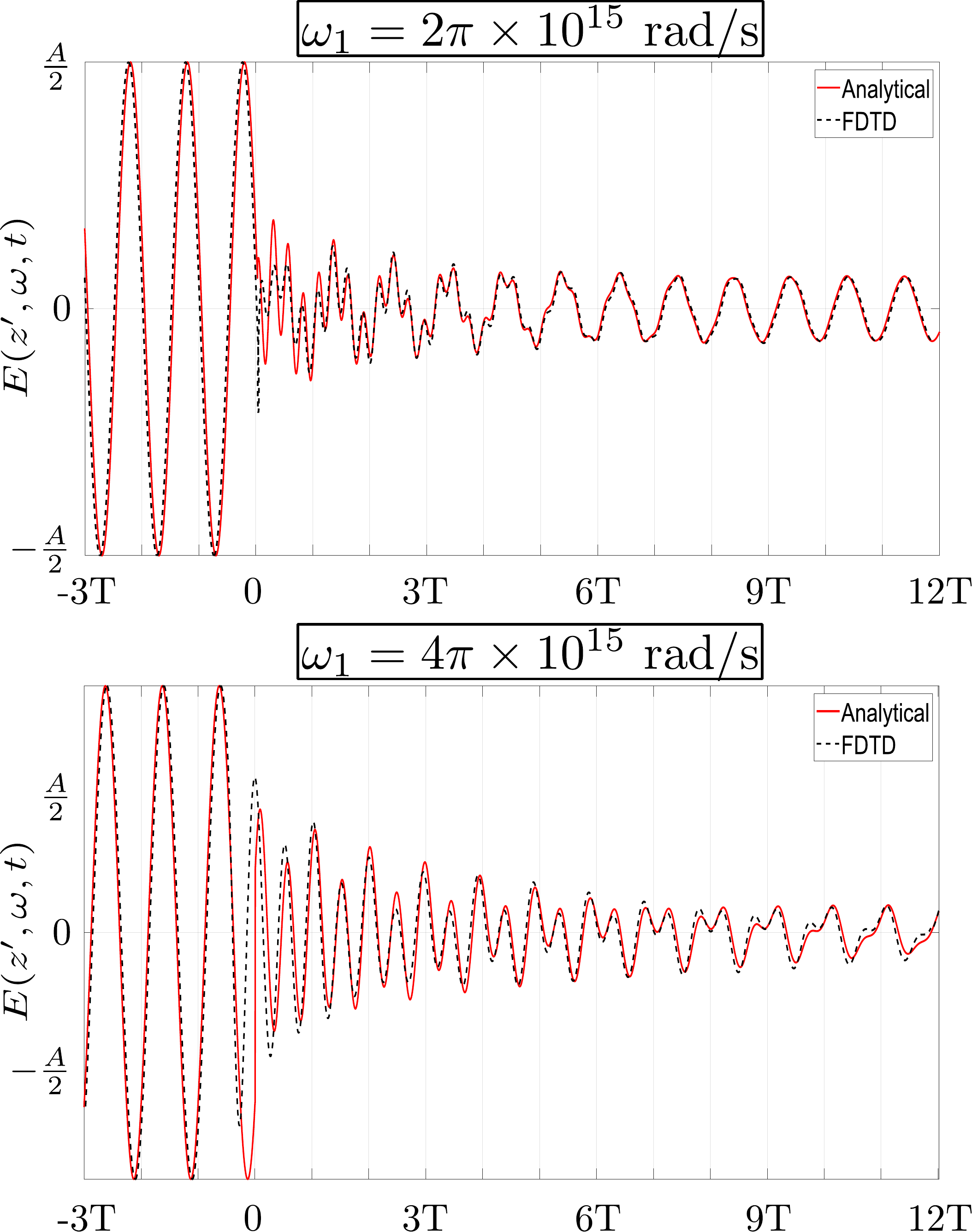}}
	\caption{Electric field evaluated at a spatial point $z'$ inside a time-varying slab that abruptly changes its properties from air to Drude at $t=0$. Realistic metals are considered for $t>0$: (a) silver, (b) aluminum. Two different frequencies of the incident wave $\omega_1$ are considered. }
	\label{Silver_aluminum}
\end{figure*}

Notice that all the analysis is carried out in the Drude-region. Both the time and spatial refractive fields are evaluated inside the metal region. Outside, the medium is always air, and the effect of the transient field is barely manifested. Since there is no time interface there, no frequency conversion can be induced in the region. Otherwise, it is true that the time-refracted forward and backward waves may reach the spatial interface and cross from Drude to air. In most cases, these fields are much less powerful than the impinging and reflected waves associated with the original field.

\section{Analytical computation. Temporal transition from air to Drude}

In order to check the validity of our previous assumptions, we will now consider a plane wave impinging on a slab that changes from air to Drude-like material at a certain instant ($t = 0$). From this instant on ($t>0$), the slab keeps being a Drude-like medium perpetually.

For simplicity, the model will be first evaluated in two real metals whose plasma frequency and damping frequency are widely known: silver and aluminum. Nevertheless, it can also be applied to various other materials whose electromagnetic response is determined by free carriers. The analytical results are corroborated with the ones provided by a self-implemented 1-D finite-difference time-domain (FDTD) framework that recreates the proposed scenario. FDTD codes are becoming the primary tool to validate theories and analytical descriptions of the physics of space-time systems \cite{vahabzadeh2017generalized, Gupta2018}. In our FDTD, a temporal step of \linebreak $\Delta t$ = 9.9 attoseconds is employed, while the computational domain is divided into 10000 cells distributed along the \textit{z}-axis and with a size of $\Delta z$ = 3 nm. The air-metal interface is placed at the center of the domain, and the open boundaries are truncated by a 32-cell perfectly matched layer (PML). The excitation is a harmonic signal whose amplitude is progressively increased at the beginning according to a Gaussian distribution in order to avoid undesired numerical high-order harmonics. The Drude material is modeled by its equivalent surface current according to the Auxiliary Differential Equation (ADE) method \cite{Taflove}. More information about the FDTD formulation can be found in the supplementary material \cite{SupplementaryMaterial}.

The first case under consideration is slab changing from air to silver at $t = 0$. Silver is characterized by its plasma frequency $\omega_p~=~1.4~\cdot~10^{16}$ rad/s and the damping constant is $\gamma~=~3.2~\cdot10^{13}\,$Hz. A plane wave with frequency $f = 10^{15}\,$Hz and amplitude $A = 0.5\,$V/m is impinging. At $t = 0$, the portion of the wave trapped in silver suffers the field transformation explained above, manifested by a time and spatial refraction. The top panel in \Fig{Silver_aluminum}(a) represents the time evolution of the field inside the slab. The time axis is normalized to the periodicity of the incident wave $T = 2\pi/\omega_1$. The field has been evaluated at a distance $z' = 19.5$ nm, from the interface, i.e., $z' = 0.065\lambda$ with $\lambda$ being the wavelength in free space. Before $t = 0$, the evolution of the field is that of the incident wave since the slab is still in air-state. After $t = 0$, the wave evolves in a different manner. The frequency has considerably increased, as expected according to the predictions given in Section II. In fact, this time evolution is a mixture of both temporal and spatial fields, with a clear dominance of the temporally-refracted backward wave. 

The scenario changes when the frequency of the incident wave is doubled [bottom panel of \Fig{Silver_aluminum}(a)]. The field is again evaluated at a distance from the interface $z' = 0.065\lambda$. Now, no temporal refraction is exhibited, i.e., beyond $t = 0$ the wave has the same frequency as before. Otherwise, there is a slight amplification in the amplitude value of the electric field. This increment comes from the fact that, at this frequency, the magnitude of the Drude relative permittivity is less than one, thus leading to a transmission coefficient greater than one in eq. \eqref{Es}. This is corroborated by introducing the silver parameters and the operation frequency in \eqref{epsilon_w}, giving $\varepsilon_{\text{r}} = -0.242 -\text{j} 0.0032\,$. The agreement with the results provided by FDTD is excellent, thus validating the theoretical and analytical framework of the problem. 

Now, we consider a scenario where the slab changes from air to aluminum at $t=0$.  The corresponding results for a Drude-like material modeling aluminum are plotted in \Fig{Silver_aluminum}(b). The damping constant is now an order of magnitude larger, $\gamma = 9.2\cdot10^{14}\,$Hz. The plasma frequency keeps the same order of magnitude, $\omega_p = 2.29\cdot 10^{16}\,$rad/s. \Fig{Silver_aluminum}(b) shows results obtained when the impinging wave has frequency $f_1 = 10^{15}\,$Hz (top panel) and $f_1 = 2\cdot10^{15}\,$Hz (bottom panel). In both cases, both the time and spatial refracted contribution are visualized from $t = 0$. In the top panel, the frequency conversion is clearly manifested up to $t = 5T$, approximately. Beyond that, the spatial refraction term governs. The bottom panel exhibits frequency conversion in a larger interval, almost reaching $t = 12T$. The fields have now been evaluated at $z' = 0.265\lambda$ in both cases, with $\lambda$ being the wavelength in free space in each case. The agreement between results from FDTD and the analytical approach is again excellent. 

It is worth mentioning that the evaluation of the temporal refraction for these cases is approximately carried out from the metal-like scenario. The behavior is more pronounced in the case of aluminum, with a field evolution manifesting a fast decay in time of its corresponding temporal refraction term. We can visualize in    \Fig{Silver_aluminum}(b) how the spatial-refraction term governs after a few periods. In \Fig{Silver_aluminum}(a) the damping constant considered is an order of magnitude lower. This means that the analysis can still be approximated under the metal-like scenario but also assuming $\gamma \ll \omega_p$. No decay is now manifested along the time interval considered. Generally, the decay exists, but it is actually slow in time in comparison to the periodicities taking part in the problem.


\begin{figure*}[!t]
	\centering
    \subfigure[]{\includegraphics[width= 0.8\columnwidth]{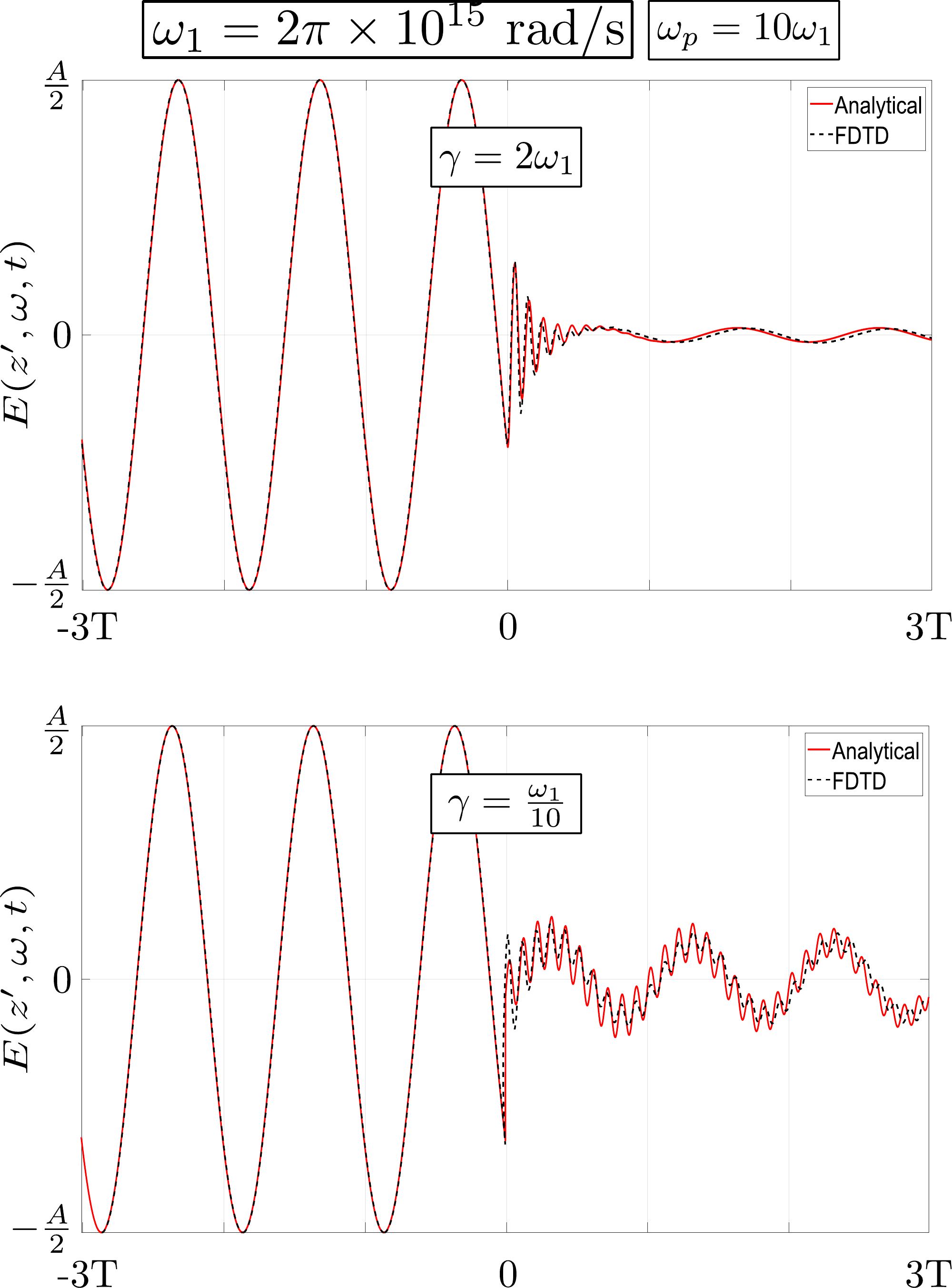}}
    \subfigure[]{\includegraphics[width= 0.8\columnwidth]{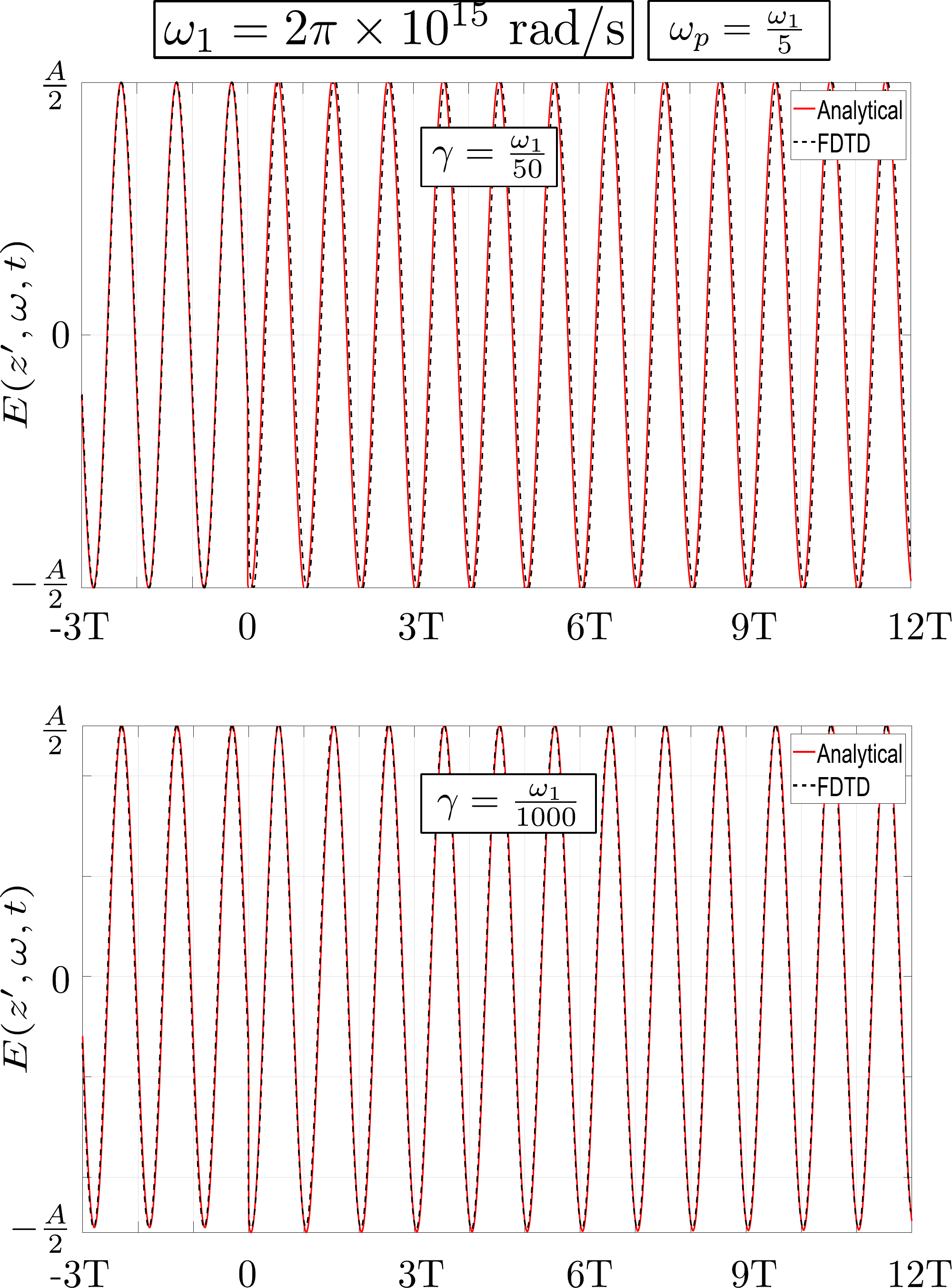}}
	\caption{Electric field evaluated at  $z'$ inside a time-varying slab that abruptly changes its properties from air to Drude at $t=0$. Different values of $\gamma$ are considered. (a) Metal-like regime ($\omega_1\leq0.1\omega_p$). (b) High-frequency (dielectric-like) regime ($\omega_1\geq5\omega_p$). }
	\label{variation_wp}
\end{figure*}

In order to check the validity of the approximate analytical solutions presented in eqs. \eqref{Et_low} and \eqref{Et_high},  a second kind of evaluation is carried out in \Fig{variation_wp}. The Drude parameters now employed do not correspond to any specific material, but they are freely modified to settle the desired scenarios. The desired scenarios regards dielectric-like scenarios and metal-like scenarios according to the definitions in Section II. The key parameter for this is the incident frequency $\omega_{1}$ (once the Drude parameters are known), although scenarios with low damping constant $\gamma$ directly enter the dielectric-like category. 

\Fig{variation_wp}(a) illustrates a scenario within the metal-like regime. A plane wave with frequency $\omega_1 = 2\pi\cdot 10^{15}\,$rad/s impinges on the interface. The top panel emulates the field in a metal slab with $\gamma = 2 \omega_{1}$ whereas the bottom panel reduces the damping constant down to $\gamma = \omega_{1}/10$. The plasma frequency is fixed to $\omega_{p} = 10\omega_{1}$,  ensuring metal-like operation. The fields are evaluated at the spatial interface $z' = 0$. In both cases, the backward contribution of the temporal refraction is visibly mixed with the direct spatial refracting field. The temporal refraction term decays rapidly over time, due to the selected large values of the damping constant. As the temporally-refracted field progressively vanishes, the spatially-refracted field starts to dominate. This is especially visible in the top panel of \Fig{variation_wp}(a), where the damping constant is so high that it takes less than one period $T$ to vanish. 

\Fig{variation_wp}(b) illustrates a second example focused on the dielectric-like regime. Now, the incident frequency $\omega_{1} = 5\omega_{\text{p}}$, is larger than the plasma frequency. In this scenario, the Drude material becomes electromagnetically transparent. According to the definitions of $E_{21}$ and $E_{22}$ in \eqref{Et_high}, it can readily be corroborated that no backward wave exists, since $\omega_{1}/\sqrt{\omega_{1}^{2} +  \omega_p^{2}} \approx 1$. This leads to $E_{21} \approx A$ (forward wave) and $E_{22} \approx 0$ (backward wave). Notice that this result does not depend on the value of $\gamma$, assumed to be negligible in this high-frequency dielectric-like regime. Therefore, the top and bottom panels of \Fig{variation_wp}(b) are almost identical,  representing the time evolution of the spatially diffracted term, which is now maximum since the metal has practically become transparent. Essentially, the incident wave does not suffer the material switching. The agreement between the numerical FDTD results and the analytical computations is good.  

Some general aspects can be inferred from the analysis of Figures \ref{Silver_aluminum} and \ref{variation_wp}. The damping frequency $\gamma$ is the main responsible for the attenuation. The higher $\gamma$ is, the quicker the temporal refraction term vanishes.   In the case of aluminum [Figure \ref{Silver_aluminum}(b)], the temporal refraction term mostly vanishes after 12  periods of the incident wave (the transient state lasts around 12 fs). Therefore,  it can be concluded from \Fig{Silver_aluminum} that the transient states generated by the temporal discontinuity at $t=0$ in elements with high $\gamma$ such as aluminum have practically no relevance at microwave and millimeter-wave frequencies. It is just when we either decrease the damping frequency $\gamma$, as in the case of using silver, or increase the operation frequency $\omega_1$, that transients become truly relevant. This can result in scenarios with richer physics such as the amplification phenomena observed in the bottom panel of Figure \ref{Silver_aluminum}(a). The results in \Fig{variation_wp} support the conclusions extracted from Figure 2. Moreover, Figure 3 also points out the fact that for very high incident frequencies $\omega_1$ transients are not relevant either. In general, it is in the intermediate region near the plasma frequency of the material ($\omega_1 \sim \omega_p$) where the transient states generated after the temporal discontinuity show the richest effects, provided that the value of $\gamma$ is adequate.

\section{Periodic Temporal Transitions}

This section is dedicated to the study of a more complex scenario. So far, a single temporal transition (from air to Drude) has been analyzed.  Now, the idea is to extend the problem to a periodic scenario where both states, air, and Drude, interchange periodically in time. Two main transitions exist in this case: from air to Drude, previously considered; and from Drude to air. The latter will be discussed in detail below.

The description of the field behavior in the temporal transition between Drude metal and the air is not trivial. Furthermore, the presence of the spatial refraction term in the slab (finite slab) introduces additional complexities to the analysis. The extension to a periodic problem therefore becomes a really complicated task. In order to simplify the analysis, let us consider that the thickness of the slab is large enough so that the spatial interfaces have no influence on the problem.

At the first task, the frequency conversion from Drude-metal to air needs to be evaluated. Assuming $\omega_{1}$ as the frequency of the original wave (in air, before any material switching) and $\omega_{3}$ as the frequency when the material becomes air again, it can be demonstrated that two solutions exist: $\omega_{1} = \pm \omega_{3}$.  This is a consequence of the wavevector continuity across all temporal interfaces, $k= k_1 = k_{2} = k_{3} $, with $k_3 = \omega_3/ c$ and $k_1, k_2$ already defined in Section II. The fact of recovering the original frequency when the material becomes air again indicates that the fields may follow a periodic tendency. As will be shown below, it is possible to induce a periodic field evolving in time with two different frequencies ($\omega_{1}$ and $\omega_{2}$), somehow facilitating frequency conversion. Notice that this mechanism of frequency conversion is different from previous examples as those in \cite{Moreno24eucap, alex2023analysis,   Moreno24spacetime}, where frequency conversion is not induced, but due to multiple-harmonics excitation (or frequency mixing). 


\begin{figure*}[!t]
	\centering
    \subfigure[]{\includegraphics[width= 0.8\columnwidth]{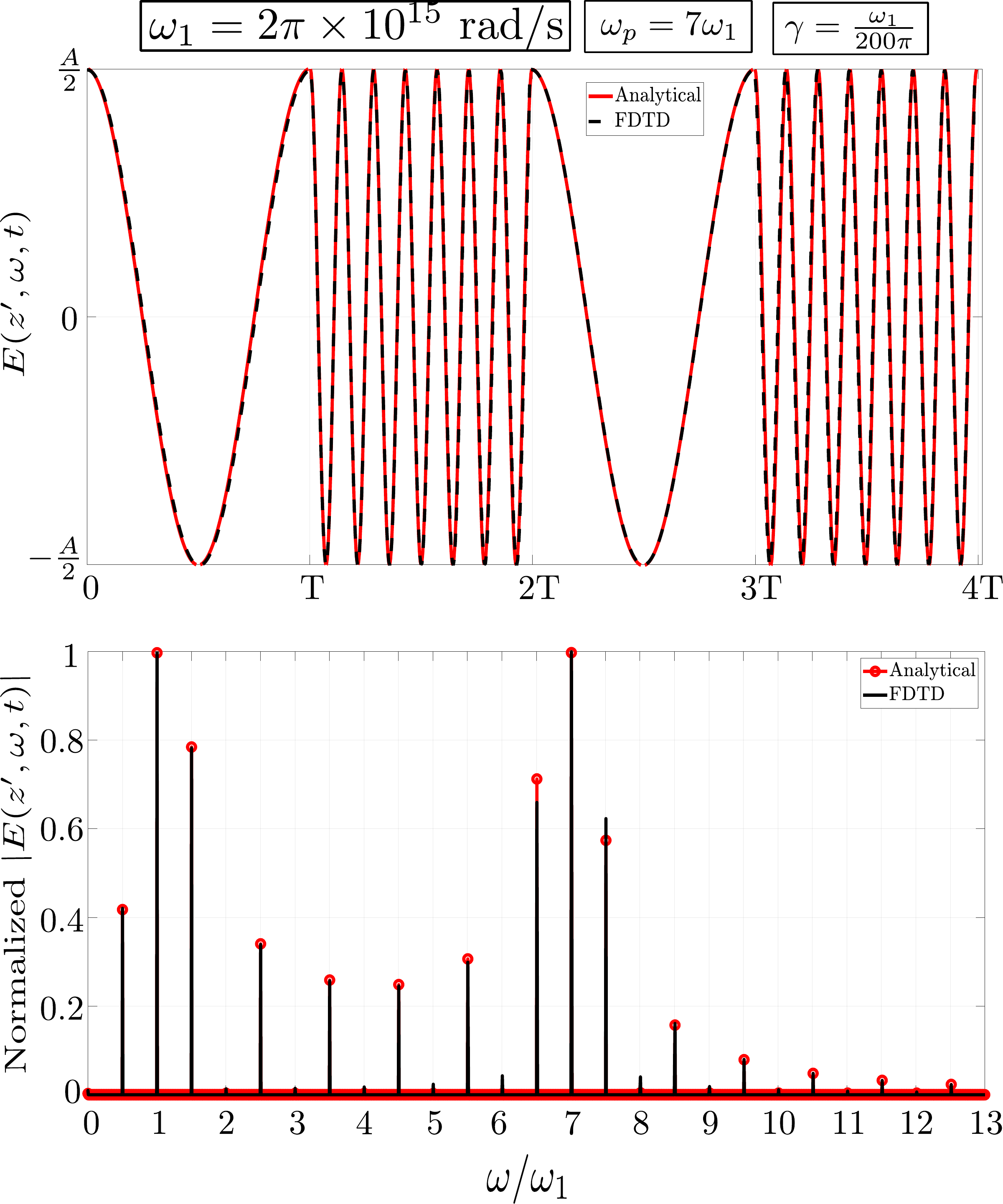}}
    \subfigure[]{\includegraphics[width= 0.8\columnwidth]{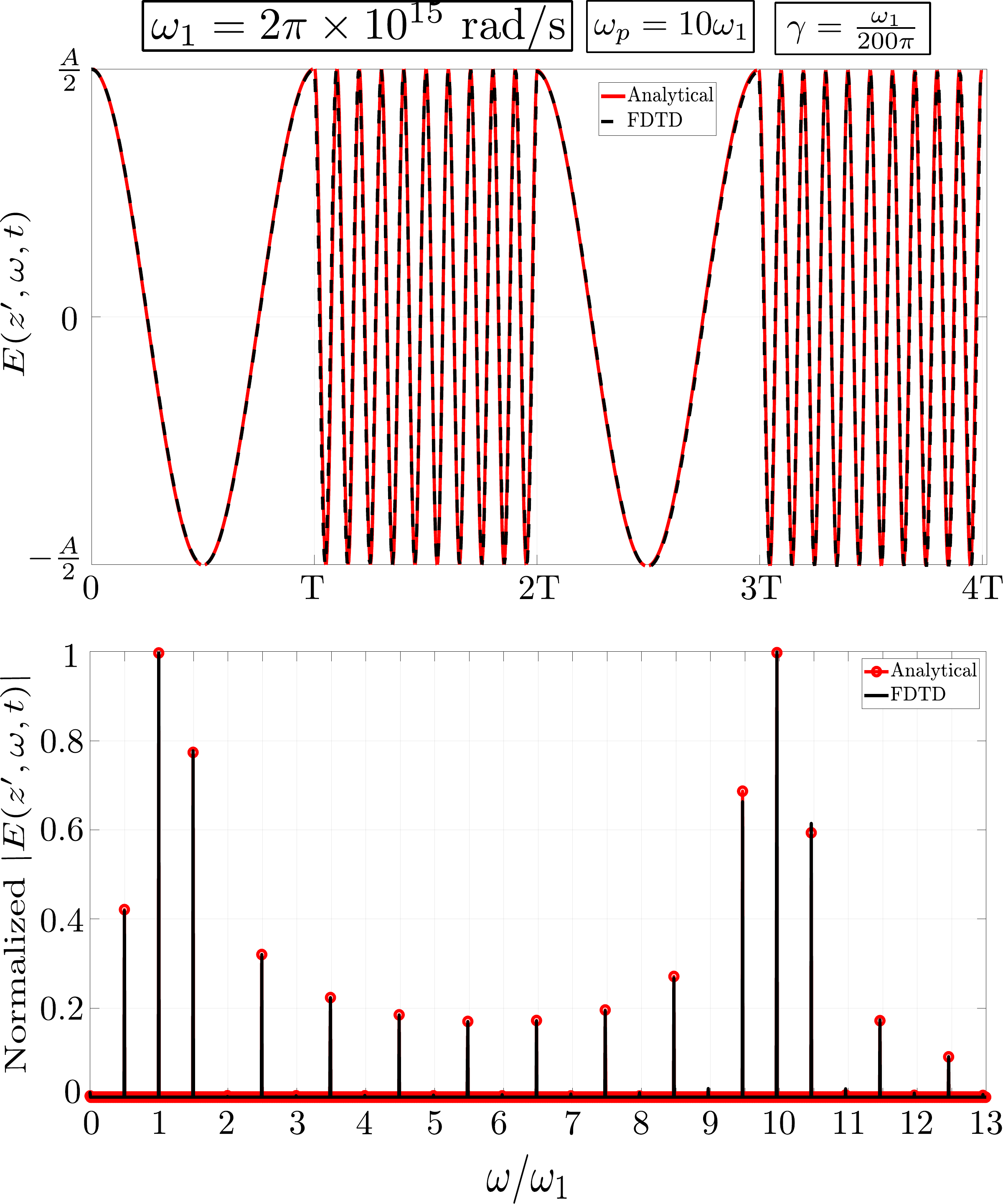}}
	\caption{Frequency conversion in a  time-varying dispersive media that periodically changes between air and Drude states. Top panels: temporal evolution of the electric field. Bottom panels: Normalized frequency spectrum. (a) $\omega_p=7\omega_1$, (b) $\omega_p=10\omega_1$.}
	\label{periodic1}
\end{figure*}

At the temporal interface, the electric-flux density vector $\mathbf{D}$ and the magnetic-flux density $\mathbf{B}$ vectors are continuous, i.e., $D(t = t^{-}, z) = D(t = t^{+}, z)$ and \linebreak $B(t = t^{-}, z) = B(t = t^{+}, z)$. The terms $t^{-}$ and $t^{+}$ refer to the instants of change just before and after the material variation from Drude ($t^{-}$) to air ($t^{+}$), respectively.  The fields at $t = t^{-}$ can be computed by using eqs. \eqref{Et_low} and \eqref{Et_high}, while the fields at $t = t^{+}$ remain unknown. Nonetheless, in the new medium (air), the electric and magnetic fields can be simply expanded as:
\begin{align}\label{D+}
D(t = t^{+}, z) &= \varepsilon_{0}\big[E_{31} \text{e}^{\text{j} \omega_{3} t^{+}} + E_{32} \text{e}^{- \text{j} \omega_{3} t^{+}} \big] \text{e}^{-\text{j} k z}  \\ 
\label{B+} B(t = t^{+}, z) &=  \frac{\omega_{3}\mu_{0} \varepsilon_{0}}{k_{3}}\big[E_{31} \text{e}^{\text{j} \omega_{3} t^{+}} - E_{32} \text{e}^{- \text{j} \omega_{3} t^{+}} \big] \text{e}^{-\text{j} k z} \, .
\end{align}
The unknowns $E_{31}$ and $E_{32}$ denote the amplitude of the forward and backward waves, respectively. Parameters $k$ and $\omega_3$ are already known after the imposition of the wavevector continuity.   

The evaluation of $D$ at $t = t^{-}$ takes into account the dispersive properties of Drude material. $D$ can be rigorously expressed as \cite{Taflove91}:
\begin{equation}\label{D-}
D(t^{-}, z) = \varepsilon_{0} E_t(t^{-}, z) + \varepsilon_{0} \displaystyle\int_{0}^{t^{-}} E_t(t^{-} - \tau) \chi (\tau) \text{d} \tau \, ,
\end{equation}
with $E_{\text{t}}(t, z)$ being the electric field in metal computed by means of the inverse Laplace transform in \eqref{E_Laplace}. Considering almost lossless Drude materials, we use the approximation $\gamma \ll \omega, \omega_p$, and $E_{\text{t}}(t^{-}, z)$ becomes that in eq. \eqref{Et_high}:
\begin{equation}\label{Et_g0}
E_{\text{t}}(t^{-}, z) \approx \frac{A}{2} \bigg[ (1 + \frac{\omega_{1}}{\omega_{2}}) \text{e}^{\jj \omega_{2}t } +(1 - \frac{\omega_{1}}{\omega_{2}}) \text{e}^{-\jj\omega_{2}t}\bigg] \text{e}^{-\text{j} k z}\, .
\end{equation}
By introducing \eqref{Et_g0} in \eqref{D-} and taking into account that $\chi(t) = \omega_p ^2 (1-\mathrm{e}^{-\gamma t}) / \gamma \stackrel{\gamma \rightarrow 0 }{\approx} \omega_p^2 t$, we  obtain
\begin{multline}\label{DDrude}
D(t^{-}, z) \approx A\varepsilon_{0}  \bigg[ \cos(\omega_{2} t^{-}) + \frac{\omega_{p}^{2}}{\omega_{2}^{2}}(1 - \cos(\omega_{2} t^{-}) ) \\ +  \text{j} \frac{\omega_{1}}{\omega_{2}} \sin(\omega_{2} t^{-})\bigg(1 - \frac{\omega_{p}^{2}}{\omega_{2}^{2}} \bigg)  \bigg]\text{e}^{-\text{j} k z}\, .
\end{multline}
In the previous expression, we have neglected the presence of a linear term proportional to $\jj \omega_1 t^{-}\ll 1$. In any case, this linear term does not manifest itself in $\mathbf{E}$ (a second time derivative is involved connecting $\mathbf{D}$ and $\mathbf{E}$ in Maxwell equations)  even if its contribution cannot be neglected. 

The corresponding $B$ in the Drude-like material is computed by applying the source-free Ampère-Maxwell equation  $\nabla~\times~\mathbf{H}(t, z)~=~\frac{\partial \mathbf{D}(t, z)}{\partial t}$ and the relation $B = \mu_0 H$, leading to
\begin{multline}\label{B-}
B(t^{-}, z) = A \frac{\mu_{0}\varepsilon_{0} }{k} \omega_{2} \bigg(1 - \frac{\omega_{p}^{2}}{\omega_{2}^{2}}\bigg) \\ \times \bigg[-\text{j}\sin(\omega_{2} t^{-}) - \frac{\omega_{1}}{\omega_{2}} \cos(\omega_{2}t^{-})  \bigg]\text{e}^{-\text{j} k z}\, .
\end{multline}

Please note that the instant of change $t^{-}$ from Drude to air can be set such that $\omega_{2} t^{-} = n 2\pi$, $n \in \mathbb{N}$. As will be shown, this is important to invoke periodic transitions between air and Drude. This particular condition reduces the expressions in \eqref{D-} and \eqref{B-} to 
\begin{align}\label{D--}
D(t^{-}, z) &\approx A\varepsilon_{0}  \text{e}^{-\text{j} k z}\\
\label{B--} B(t^{-}, z) &\approx - A \frac{\mu_{0}\varepsilon_{0}}{k} \omega_{1} \bigg(1 - \frac{\omega_{p}^{2}}{\omega_{2}^{2}} \bigg) \text{e}^{-\text{j} k z}
\end{align}
The application of the temporal boundary conditions at $t=t^-=t^+$ leads to the following electric field in the air region:
\begin{equation} \label{E_periodic}
    E(z,t) = \text{e}^{-\text{j} k z} \Big[E_{31} \text{e}^{\text{j}\omega_{3} (t - t^{+})} + E_{32} \text{e}^{-\text{j}\omega_{3} (t - t^{+})} \Big]\,
\end{equation}
with
\begin{align}\label{E+}
E_{31} &= \frac{A}{2}  \left[1 + \left(1 - \frac{\omega_p^{2}}{\omega_{2}^{2}}\right)\right]\\
\label{E-}
E_{32} &= \frac{A}{2}  \left[1 - \left(1 - \frac{\omega_p^{2}}{\omega_{2}^{2}}\right)\right]\, .
\end{align}

Notice that, for a given frequency of the incident wave $\omega_{1} \le 0.14 \omega_p$, the factor $\omega_p/\omega_{2} \ge 0.99$, leading to \linebreak  $(1 - \frac{\omega_p^{2}}{\omega_{2}^{2}}) \le 0.01$. Similar conditions can be found even at higher frequencies. In most cases, the electric-field components $E_{31} = E_{32} \approx A/2$. This fact states that the original wave can be recovered under certain conditions, especially when $\omega_{3} (t - t^{+}) = 2 n \pi$, where $E(z, t)$ in eq. \eqref{E_periodic} simplifies to $E(z, t)~=~A\text{e}^{-\text{j} k z}$. This result opens up the possibility of creating permanent periodic temporal conditions by conveniently switching metal and Drude states. In fact, it can be demonstrated that when the switching instant coincides with a full cycle of the waves in both air and Drude  ($\omega_{1} t = \omega_{2} t = 2n\pi$), the field behavior at a given $z$ point becomes periodic.  

The above statements are tested in \Fig{periodic1}, where a thick time-varying slab periodically alternates between air and Drude states. In both cases, the frequency of the original wave is $\omega_{1} = 2\pi \cdot 10^{15}\,$rad/s and the damping constant is $\gamma =  10^{13}\,$Hz. Note that the damping constant is far from being null, although it is significantly smaller than $\omega_1$ and $\omega_p$. This choice of $\gamma$ ensures an almost lossless dielectric-like regime, especially when fast material variations are considered. The case in \Fig{periodic1}(a) denotes a Drude material with $\omega_p = 7 \omega_{1}$. The material changes periodically with $T = 2\pi/\omega_1$. The selected period results in one wave cycle in air, and seven wave cycles in the Drude. The instant of change always coincides with a maximum of the electric field. The time field variation is as expected, exhibiting a $2T$-periodic field. The Fourier transform of this field gives information about the harmonics taking part. The bottom panel in \Fig{periodic1}(a) represents the frequency spectrum of the periodic signal. In the spectrum, two harmonics stand out for having a greater relevance: first one at $\omega/\omega_{1} = 1$, coinciding with the frequency of the wave in the air; second one at $\omega/\omega_{1} = 7$, which coincides with the frequency of the wave in the Drude medium.  As observed, frequency mixing is induced as a consequence of the temporal change in the electrical properties of the materials. 

The top panel in \Fig{periodic1}(b) illustrates the temporal evolution of the electric field when $\omega_p = 10\omega_{1}$ (the rest of the parameters are kept the same as in Figure \ref{periodic1}(a)). A full period, $2T$, includes a single cycle in air and ten cycles in the Drude. In the frequency spectrum showed in the bottom panel of  \Fig{periodic1}(b), two harmonics carry the greatest amplitude: a first one, created by the wave propagating in air, located at $\omega/\omega_1 = 1$; and a second one, created by the wave propagating in the Drude medium, located at $\omega/\omega_1 = 10$. The agreement with the results provided by FDTD is excellent in all cases.

Figures \ref{Silver_aluminum}-\ref{periodic1} have shown that frequency mixing, power transfer between harmonics, and field amplification/attenuation can be tailored by artificially switching over time the parameters of a certain material or, equivalently, the homogenized effective parameters of a metamaterial. As mentioned in the previous sections, the parameters $\omega_p$ and $\gamma$ can be externally tuned in the laboratory in some Drude-like materials. In addition, the illumination frequency $\omega_1$ can significantly modify the physical response of the spatiotemporal system. More importantly, the transient responses of the electromagnetic fields generated after each temporal switching should not be perceived as design constraints or unavoidable phenomena. Transients take part of the whole mechanism to induce efficient frequency conversion and field amplification. When used conveniently, transient states have the potential to open up new possibilities and mechanisms to achieve unconventional electromagnetic and optical responses. 


\section{Conclusion}

This paper has presented an analytical framework for the study of transients in dispersive spatiotemporal slabs. Frequency dispersion is taken into consideration with the Drude model. The electromagnetic fields can be approximately described with two terms: one coming from temporal refractions and the other associated with spatial refractions. Analytical solutions are obtained in the metal-like ($\omega_1\ll \gamma, \omega_p$) and dielectric-like regimes ($\gamma \ll  \omega_1$). The continuity of the electromagnetic fields, wavevectors, and frequencies at the space-time boundaries is also discussed.  Results, validated with a self-implemented FDTD code, include the analysis of the formation of transients in time-varying metallic slabs (silver, aluminum), parameter sweeps on the plasma and damping frequencies, and a final section recreating a temporally-periodic system formed by the alternation of air and Drude states. Results show the proper manipulation of the transient states can shape the overall electromagnetic response of the time-varying system, leading to frequency conversion phenomena and field amplification/attenuation, which is of potential interest in engineering applications.

The reader is referred to the supplementary material \cite{SupplementaryMaterial} for a deeper explanation regarding the considered approximations when establishing the high(low)-frequency regimes examined in Section \ref{sec:temporal_refraction_term}. Moreover, the frequency limits for the presented formulation are included, as well as a description of the employed FDTD scheme, that implements time-varying capabilities and material dispersion.

\section*{Acknowledgement}

This work was supported in part by the Grant No. IJC2020- 043599-I funded by MICIU/AEI/10.13039/501100011033 and by European Union NextGenerationEU/PRTR, in part by the Spanish Government under Project PID2020-112545RB-C54, PDC2022-133900-I00, PDC2023-145862-I00, TED2021-129938B-I00 and TED2021-131699B-I00.

\bibliography{ref}



\renewcommand{\figurename}{Figure}






\setcounter{equation}{0}
\setcounter{figure}{0}
\setcounter{table}{0}
\setcounter{page}{1}
\makeatletter
\renewcommand{\theequation}{S\arabic{equation}}
\renewcommand{\thefigure}{S\arabic{figure}}


\widetext
\begin{center}
\textrm{\huge \textbf{Supplementary Material}}
\end{center}

\vspace{0.2cm}

\begin{center}
\textrm{\Large Transient States to Control the Electromagnetic Response of\\ Space-time Dispersive Media}
\end{center}

\begin{center}
\textrm{\large Pablo H. Zapata-Cano$^1$, Salvador Moreno-Rodríguez$^2$, Stamatios Amanatiadis$^1$, \\ Antonio Alex-Amor$^{3}$, Zaharias D. Zaharis$^{1}$, Carlos Molero$^{2}$}
\end{center}

\begin{center}
\textit{\large \vspace{0.1cm}$^1$School of Electrical and Computer Engineering, Aristotle University of Thessaloniki, \\ 54124 Thessaloniki, Greece \vspace{0.1cm}\\ 
$^2$Department of Signal Theory, Telematics and Communications, Research Centre for Information and Communication Technologies (CITIC-UGR), Universidad de Granada, 18071 Granada, Spain \vspace{0.1cm}\\ 
$^{3}$Department of Electrical and Systems Engineering, University of Pennsylvania,  Philadelphia, Pennsylvania 19104, United States}
\end{center}

\section*{Frequency Pumping in a Thick Drude Slab}

\subsection*{Air-to-Drude Transition}

If we consider a sufficiently thick slab where the spatial boundaries have no influence, the spatiotemporal problem can be reduced to a purely temporal one. In this limit, the wavenumber must be continuous after the temporal jump at $t=0$. By defining  $k_1 = k(t = 0^{-})$ and $k_2 = k(t = 0^{+})$, an air-to-Drude temporal transition must imply
\begin{equation}
    k_1 = \frac{\omega_1}{c} =  k_2 = \frac{\omega_2}{c} \sqrt{1 - \frac{\omega_p^2}{\omega_2^2 - j\omega_2\gamma }}\, .
\end{equation}
A proper manipulation of the former equation leads to a complex-valued cubic equation for the new frequency $\omega_2$:
\begin{equation}
    \omega_2^3 -\jj\omega_2^2\gamma - \omega_2(\omega_p^2+\omega_1^2)+\jj\omega_1^2\gamma = 0\, ,
\end{equation}
which, in a general situation, possesses three complex-valued solutions.

Analytical solutions are quite tedious in the general case and can obscure the understanding of the underlying physical phenomena. However, simplified and physically insightful analytical solutions can be obtained in the low-frequency and high-frequency regimes. In the limit of the low-frequency regime, $\omega_1 \rightarrow 0$, and the analytical expressions for the new frequencies $\omega_2$ take the form
\begin{equation} \label{omega_low}
    \omega_2 \approx \frac{1}{2} \left(\pm \sqrt{4\omega_p^2 - \gamma^2} +j\gamma \right), \quad \text{low freq.}
\end{equation}
For most metals that are good conductors, it is true that $\omega_p \gg \gamma$, so $\pm \sqrt{4\omega_p^2 - \gamma^2} \approx \pm 2\omega_p$. In these cases, eq. \eqref{omega_low} can be further simplified to $\omega_2 \approx \pm \omega_p + j\gamma / 2$. From a practical computational perspective, the low-frequency approximation has proved to give accurate results as long as $\omega_1 < 0.1 \omega_p$ (tens or even hundreds of THz, considering the standard values of $\omega_p$ in good-conducting metals). 

At high frequencies, the Drude material becomes electromagnetically transparent, so it can be considered that $\gamma \rightarrow 0$. In this scenario, the new frequencies $\omega_2$ take the form
\begin{equation} \label{omega_high}
    \omega_2 \approx \pm \sqrt{\omega_1^2 + \omega_p^2}, \quad \text{high freq.}
\end{equation}
%

\subsection*{Drude-to-Air Transition}

In a Drude-to-air transition, the conservation of the wavenumber implies that the new frequency $\omega_3$ in the air medium is directly extracted from the frequency $\omega_2$ in the Drude material, that is,
\begin{equation} \label{omega3}
    \frac{\omega_3}{c} = \frac{\omega_2}{c} \sqrt{1 - \frac{\omega_p^2}{\omega_2^2 - j\omega_2\gamma }}\, .
\end{equation}

It can be analytically shown, that the original frequency $\omega_1$ is recovered in the air; namely, $\omega_3 = \omega_1$. In the low-frequency approximation ($\omega_1 \rightarrow 0$), the frequency $\omega_2$ was estimated as $\omega_2 \approx \frac{1}{2} \left(\pm \sqrt{4\omega_p^2 - \gamma^2} +\jj\gamma \right)$. By inserting $\omega_2$ in eq. \eqref{omega3}, we obtain that 
\begin{multline}
    \omega_3 \approx \left[\frac{1}{2} \left(\pm \sqrt{4\omega_p^2 - \gamma^2} +\jj\gamma \right)\right] \sqrt{1 - \frac{\omega_p^2}{\frac{1}{4} \left(\pm \sqrt{4\omega_p^2 - \gamma^2} +\jj\gamma \right)^2  \mp \jj \gamma \frac{1}{2} \left(\pm \sqrt{4\omega_p^2 - \gamma^2} +\jj\gamma \right)} } \\
    \stackrel{\gamma \ll \omega_p }{\approx}
    \left[\frac{1}{2} \left(\pm \sqrt{4\omega_p^2 - \gamma^2} +\jj\gamma \right)\right] \sqrt{1 - \frac{\omega_p^2}{\omega_p^2}} \approx 0
\end{multline}
Thus, $\omega_3 = \omega_1 \approx 0$ in the low-frequency scenario.

In the high-frequency regime ($\gamma \rightarrow 0$), the frequency $\omega_2$ was estimated as $\omega_2 \approx \sqrt{\omega_p^2 + \omega_1^2}$. By inserting $\omega_2$ in eq. \eqref{omega3}, we obtain that
\begin{equation}
    \omega_3 \approx \pm \sqrt{\omega_p^2 + \omega_1^2}\, \sqrt{1 - \frac{\omega_p^2}{\omega_p^2 + \omega_1^2}} = \pm \sqrt{\omega_p^2 + \omega_1^2}\, \sqrt{ \frac{\omega_1^2}{\omega_p^2 + \omega_1^2}} = \pm \omega_1\, .
\end{equation} 

\section*{On the Metal-like and Dielectric-like Regimes}

The relative permittivity $\varepsilon(\omega)$ associated with the Drude model is given by eq. (1) of the manuscript. It is a  complex-valued term that can be divided into its real and imaginary parts  as $\varepsilon(\omega) = \varepsilon'(\omega) + \jj\varepsilon''(\omega) $, with
\begin{equation} \label{eps_prime}
    \varepsilon'(\omega) = 1 - \frac{\omega_p^2}{\omega^2 + \gamma^2}
\end{equation}
\begin{equation} \label{eps_twoprime}
\varepsilon''(\omega) = -\frac{\gamma \omega_p^2}{\omega^3 + \omega \gamma^2}
\end{equation}

\subsection*{Frequency Limit for the Metal-like Regime}

The condition used in the manuscript to derive the formulation associated with the metal-like regime is that $\omega_1 \rightarrow 0$ (the incident frequency is zero). At low frequencies, the Drude material should behave like a good-conducting metal, reflecting most of the incident wave and rapidly absorbing the tiny part that enters the material. This behavior is associated with large values for the real and imaginary parts of the Drude permittivity, the real part being negative. A value $-\omega_p^2 / (\omega_1^2 + \gamma^2) = -50$ could be considered large enough from a practical perspective. This leads to $\omega_1 < \sqrt{\omega_p^2 / 50 - \gamma^2} \approx \omega_p/7$, provided that $\omega_p \gg \gamma$. We could even reduce further this limit to $\omega_1<0.1\omega_p$ to be conservative. For $\omega_1=0.1\omega_p$, the real part of the permittivity is $\varepsilon' \approx -99$, while the imaginary part is $\varepsilon'' \approx 1000\gamma / \omega_p$. From DC to $\omega_1 = 0.1\omega_p$, the metal-like approximation holds.   

\subsection*{Frequency Limit for the Dielectric-like Regime}

The condition used in the manuscript to derive the analytical dielectric-like regime is that $\gamma \rightarrow 0$. The former condition implies that the Drude material behaves as a dielectric with a small, but not necessarily negligible, loss term. This is the situation when $\omega_1^2 \gg \omega_1 \gamma$  or, equivalently, $\omega_1 \gg \gamma $. Nonetheless, unlike in metal-like and high-frequency regimes, it is not that easy here to estimate a lower bound for $\omega_1$. This is because the imaginary part of the Drude permittivity $\varepsilon''$ is actually dependent on the considered plasma frequency $\omega_p$ [see eq. \eqref{eps_twoprime}]. Therefore, each case should be treated individually, provided always that $\omega \gg \gamma$. However, in the particular case of dealing with good conductors such as copper, silver, aluminum, or gold ($\omega \sim 10^{16}, \gamma \sim 10^{14}$) characterized via the Drude model, the dielectric-like regime approximation could be perfectly applied as long as $\omega_1 \geq 5\cdot 10^{15}$ rad/s.

\subsection*{Frequency Limit for the High-Frequency Regime}

The high-frequency regime should be considered as a special case within the dielectric-like regime. In the high-frequency approximation, the Drude material becomes electromagnetically transparent, i.e., it behaves like air. To meet this condition, we need that $\varepsilon' \approx 1$ and that $\varepsilon'' \approx 0$. by inspecting eqs. \eqref{eps_prime} and \eqref{eps_twoprime}, it can be readily inferred that the high-frequency regime occurs when $\omega_1 \gg \omega_p$.  The condition $\omega_1 \geq 5\omega_p$ establishes a conservative limit beyond which the high-frequency approximation holds. This limit guarantees that $0.96\leq \varepsilon' \leq 1 $ and $|\varepsilon''| \lesssim \gamma/(125\omega_p) \ll 1$.

\subsection*{State of the Art on $\omega_p$ and $\gamma$ Modulation}
Several prototypes that modify the carrier density ($N$) and carrier mobility ($\mu_e$) are found in the literature. These parameters affect directly to $\omega_p$ and $\gamma$ following next equations:
\begin{equation} \label{w_p}
\omega_p = \sqrt{\frac{N \cdot q_{e}^2}{\varepsilon_0 \cdot m_e }}\, ,
\end{equation}
\begin{equation} \label{gamma}
\gamma = \frac{q_e}{\mu_e \cdot m_e }\, ,
\end{equation}
where $q_e$ represents the electron charge, $\varepsilon_0$ is the electric permittivity in the vacuum and $m_e$ is the electron effective mass. TABLE \ref{table1} shows several references relative to the modification of $N$ under external electric stimulation and $\mu_e$ applying strain engineering.

\begin{table}[!h]
\caption{Variability of the carrier density ($\Delta N$) and carrier mobility ($\Delta\mu_e$) as a function of external stimulation or strain techniques. These parameters affect directly and independently the plasma frequency ($\omega_p$) and damping frequency ($\gamma$).}
\centering \label{table1}
\begin{tabular}{c *6c} 
\toprule
$\Delta_{EXT}$ & Spectrum & Material & $\Delta_N$  & $\Delta_{\mu_e}$ & Ref.\\ 
\toprule
$2.5$V & $800$nm & ITO & $\approx2.7\cdot10^{22}$ cm$^{-3}$  &  N.A. &  $[37]$ \\
\midrule
$5$V & $1500$nm & ITO & $\approx5.5\cdot10^{20}$ cm$^{-3}$  &  N.A. &  $[38]$  \\
\midrule
$2.5$V & $532$nm & AZO & $\approx72\cdot10^{20}$ cm$^{-3}$ &  N.A. & $[39]$ \\
\midrule
Strain of $5\%$ & N.A. & WSe$_2$ & N.A. &  $\approx80$ cm$^2$/Vs & $[44]$  \\
\midrule
Strain of $4\%$  & N.A. &2L r-MoS$_2$ & N.A. &  $\approx490$ cm$^2$/Vs &  $[45]$  \\
\midrule
 \end{tabular}

\end{table}

\section*{FDTD model for time-varying media}
\begin{figure*}[!h]
	\centering
    \subfigure{\includegraphics[width= 0.3\columnwidth]{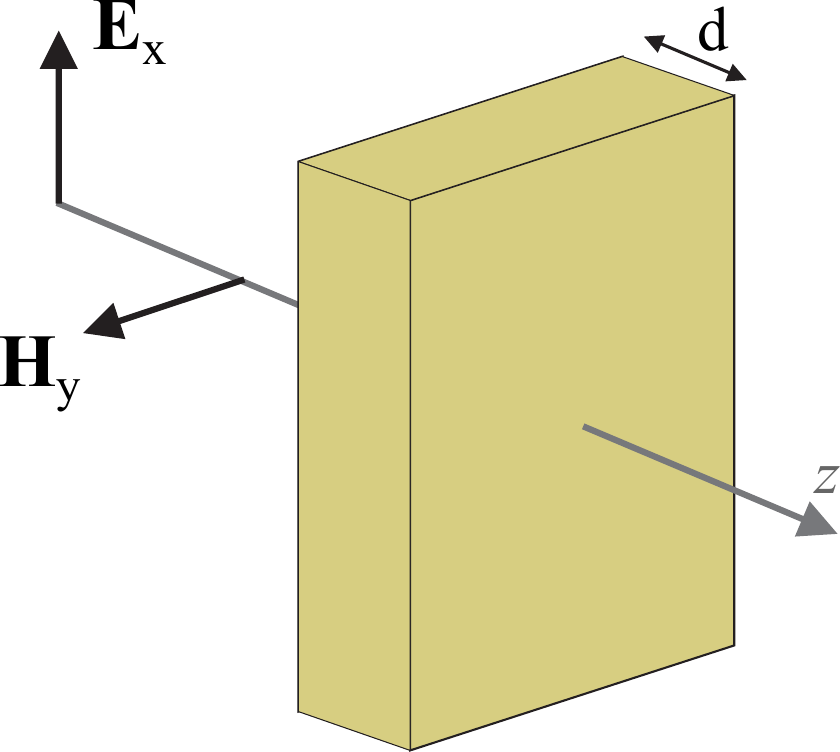}}    
    \caption{Schematic of the one-dimensional Yee-cell employed in the FDTD algorithm. }
	\label{yeeCell}
\end{figure*}

Figure \ref{yeeCell} depicts the Yee-cell employed in our scheme. As it can be observed, it is a 1D problem (normal incidence is considered in all scenarios) in which the yellow block represents the Drude material slab of thickness \textit{d} placed on the x-y plane.

According to this notation, the update equation for the electric field in the main loop of the FDTD algorithm is given by

\begin{equation}
    {\mathbf{E}}_x|^{n+1}_{k+\frac{1}{2}}=C_a|^{n+1}_{k+\frac{1}{2}}\cdot  {\mathbf{E}}_x|^{n}_{k+\frac{1}{2}} + C_   b|^{n+1}_{k+\frac{1}{2}} \left ( \frac{{\mathbf{H}}_y|^{n+\frac{1}{2}}_{k+1} - {\mathbf{H}}_y|^{n+\frac{1}{2}}_{k}}{\Delta z} + \frac{(1+k_p)\cdot {\mathbf{J}}_{d}|^{n+\frac{1}{2}}_{k+\frac{1}{2}}}{2} \right )
    \label{eq:updateE}
\end{equation}

where $n$ indicates the current time step ($t_n= n\Delta t$, being $\Delta t$  the time step considered for the simulation) and $k$ is an integer that corresponds to the z-coordinate ($\Delta z$ stands for the cell size). An \textit{x}-polarized wave traveling along the \textit{z}-axis is considered. According to the Auxiliary Differential Equation (ADE) method, the Drude slab is modeled by its equivalent surface current, which takes the following form:

\begin{equation}
{\mathbf{J}}_{d}|^{n+\frac{1}{2}}_{k+\frac{1}{2}} =
\begin{cases}
        k_p \cdot {\mathbf{J}}_{d}|^{n-\frac{1}{2}}_{k+\frac{1}{2}} + b_p \left ( {\mathbf{E}}_x|^{n+1}_{k+\frac{1}{2}} + {\mathbf{E}}_x|^{n}_{k+\frac{1}{2}} \right ),  & \text{if "Drude" state at time } n\Delta t \\
      0, & \text{if "air" state at time } n\Delta t
\end{cases}
\label{eq:CurrentJ}
\end{equation}
The coefficients $k_p$ and $b_p$ depend on the Drude properties of the material (i.e. the plasma frequency $\omega_p$ and the damping constant $\gamma$ ) and are calculated as

\begin{equation}
    k_p = \frac{1-\gamma \Delta t/2}{1+\gamma \Delta t/2} \ , \quad b_p = \frac{\omega_p^2 \varepsilon_0 \Delta t/2}{1+\gamma \Delta t/2} 
\end{equation}

$C_a$ and $C_b$ are material parameters that take the following forms for the "air" and "Drude" states:

\begin{equation}
 C_a|^{n}_{k+\frac{1}{2}}=C_a(n\Delta t) =
\begin{cases}
           \dfrac{2 \varepsilon_0 -b_p \Delta t }{2 \varepsilon_0+b_p\Delta t  },  & \text{if "Drude" state at time } n\Delta t \\
      \hfil 1, & \text{if "air" state at time } n\Delta t
\end{cases}
\end{equation}

\begin{equation}
 C_b|^{n}_{k+\frac{1}{2}}=C_b(n\Delta t) = 
\begin{cases}
           \dfrac{2 \Delta t}{2 \varepsilon_0 +b_p\Delta t },  & \text{if "Drude" state at time } n\Delta t \vspace{2mm}\\
      \hfil \dfrac{\Delta t}{\varepsilon_0}, & \text{if "air" state at time } n\Delta t
\end{cases}
\end{equation}

Thus, the ADE-FDTD technique employed for modeling the Drude material can be seen as a three-step procedure. Starting with the stored components of $\mathbf{E}^n$, $\mathbf{H}^n$, and $\mathbf{J}^n$ (from the previous iteration), $\mathbf{E}^{n+1}$ is calculated using \ref{eq:updateE}. Then, the value of the just-computed $\mathbf{E}^{n+1}$ is inserted into \ref{eq:CurrentJ}. Finally, the value of $\mathbf{H}^{n+1}$ is obtained in the classic manner by the Yee realization of Faraday's law (see Taflove 2005, ref [50] in the manuscript), completing thus the cycle that starts again for the next iteration.



\end{document}